\documentclass[journal=jpcafh]{achemso}

\usepackage[version=3]{mhchem} % Formula subscripts using \ce{}
\usepackage{fontenc}
\usepackage{chemformula}
\usepackage{amsmath}
\usepackage{amsfonts}
\usepackage{amssymb}
\usepackage{lineno}
\usepackage{mathtools}
\usepackage{xcolor}
\usepackage{setspace}
\usepackage{pdfpages}

\usepackage{color}
\author{Federico Mazzola}
\affiliation{CNR-IOM TASC Laboratory, Area Science Park, I-34149 Trieste, Italy}
\affiliation{Department of Molecular Sciences and Nanosystems, Ca Foscari University of Venice, 30172 Venice, Italy}
\email{federico.mazzola@unive.it}

\author{Barun Ghosh}
\affiliation{Department of Physics, Northeastern University, Boston, Massachusetts 02115, USA}
\email{barunghosh02@gmail.com}

\author{Jun Fujii}
\affiliation{CNR-IOM TASC Laboratory, Area Science Park, I-34149 Trieste, Italy}

\author{Gokul Acharya}
\affiliation{Department of Physics, University of Arkansas, Fayetteville, AR 72701}

\author{Debashis Mondal}
\affiliation{CNR-IOM TASC Laboratory, Area Science Park, I-34149 Trieste, Italy}

\author{Giorgio Rossi}
\affiliation{University of Milano, I-20133 Milano, Italy}
\affiliation{CNR-IOM TASC Laboratory, Area Science Park, I-34149 Trieste, Italy}

\author{Arun Bansil}
\affiliation{Department of Physics, Northeastern University, Boston, Massachusetts 02115, USA}

\author{Daniel Farias}
\affiliation{Departamento de Física de la Materia Condensada, Universidad Autónoma de Madrid, 28049, Madrid, Spain}
\affiliation{Instituto "Nicolas Cabrera" and Condensed Matter Physics Center (IFIMAC), Universidad Autonoma de Madrid, 28049, Madrid, Spain}

\author{Jin Hu}
\affiliation{Department of Physics, University of Arkansas, Fayetteville, AR 72701}

\author{Amit Agarwal}
\affiliation{Department of Physics, Indian Institute of Technology Kanpur, Kanpur 208016, India}

\author{Antonio Politano}
\affiliation{Department of Physical and Chemical Sciences,
University of L' Aquila, via Vetoio 67100 L' Aquila (AQ), Italy}

\author{Ivana Vobornik}
\affiliation{CNR-IOM TASC Laboratory, Area Science Park, I-34149 Trieste, Italy}

\title{Discovery of a magnetic Dirac system with large intrinsic non-linear Hall effect}

\abbreviations{ARPES, DFT}

\begin{document}

%%%%%%%%%%%%%%%%%%%%%%%%%%%%%%%%%%%%%%%%%%%%%%%%%%%%%%%%%%%%%%%%%%%%%
%% The "tocentry" environment can be used to create an entry for the
%% graphical table of contents. It is given here as some journals
%% require that it is printed as part of the abstract page. It will
%% be automatically moved as appropriate.
%%%%%%%%%%%%%%%%%%%%%%%%%%%%%%%%%%%%%%%%%%%%%%%%%%%%%%%%%%%%%%%%%%%%%

%%%%%%%%%%%%%%%%%%%%%%%%%%%%%%%%%%%%%%%%%%%%%%%%%%%%%%%%%%%%%%%%%%%%%
%% The abstract environment will automatically gobble the contents
%% if an abstract is not used by the target journal.
%%%%%%%%%%%%%%%%%%%%%%%%%%%%%%%%%%%%%%%%%%%%%%%%%%%%%%%%%%%%%%%%%%%%%

\newpage

\begin{abstract}
Magnetic materials exhibiting topological Dirac fermions are attracting significant attention for their promising technological potential in spintronics. In these systems, the combined effect of the spin-orbit coupling and magnetic order enables the realization of novel topological phases with exotic transport properties, including the anomalous Hall effect and magneto-chiral phenomena. Herein, we report experimental signature of topological Dirac antiferromagnetism in TaCoTe$_2$ via angle-resolved photoelectron spectroscopy (ARPES) and first-principles density functional theory (DFT) calculations. In particular, we find the existence of spin-orbit coupling-induced gaps at the Fermi level, consistent with the manifestation of a large intrinsic non-linear Hall conductivity. Remarkably, we find that the latter is extremely sensitive to the orientation of the Néel vector, suggesting TaCoTe$_2$ a suitable candidate for the realization of non-volatile spintronic devices with an unprecedented level of intrinsic tunability.
\end{abstract}

%%%%%%%%%%%%%%%%%%%%%%%%%%%%%%%%%%%%%%%%%%%%%%%%%%%%%%%%%%%%%%%%%%%%%
%% Start the main part of the manuscript here.
%%%%%%%%%%%%%%%%%%%%%%%%%%%%%%%%%%%%%%%%%%%%%%%%%%%%%%%%%%%%%%%%%%%%%

\newpage

Over the last few years, magnetic systems with Dirac-like electronic dispersion \cite{Young_2017,Zhang_2016, Tokura_2019,Pierantozzi_2022} have been under the spotlight of both theoretical and experimental investigations,  since they are expected to support several exotic transport phenomena, including the anomalous and non-linear Hall effects \cite{Liu_2018, Gao_2021, Liu_2019}. In particular, the combined action of magnetic order and spin-orbit coupling (SOC) results in the opening of energy gaps in the electronic spectrum, within which non-trivial topological phases can occur \cite{Sekine_2016, Rashba_2009, Bernevig_2006, Yasuhiro_2015, Bruno_2004}. Such phases, in the presence of long-range magnetic order, can be modulated with ease, thus offering a methodology to tune the topological protection, hence the charges and spins \cite{Pierantozzi_2022}. Here, we report the experimental discovery of the new system TaCoTe$_2$, a topological antiferromagnet with Dirac-like dispersion and SOC-induced gaps at the Fermi level \cite{Fukami_2020, Baltz_2018, Li_2019}. We discover that the combined effect of SOC and magnetic order enables the realization of a large intrinsic non-linear Hall effect (INHE) which can be tuned by magnetic fields. Our discovery suggests TaCoTe$_2$ as a promising candidate for the realization of highly-controllable topological devices that will exploit SOC-derived transport phenomena \cite{Du_2021, Murakami_2006, Wunderlich_2005, Sun_2016}.

\begin{figure}[ht]
\centering
    \includegraphics[width=0.45\textwidth]{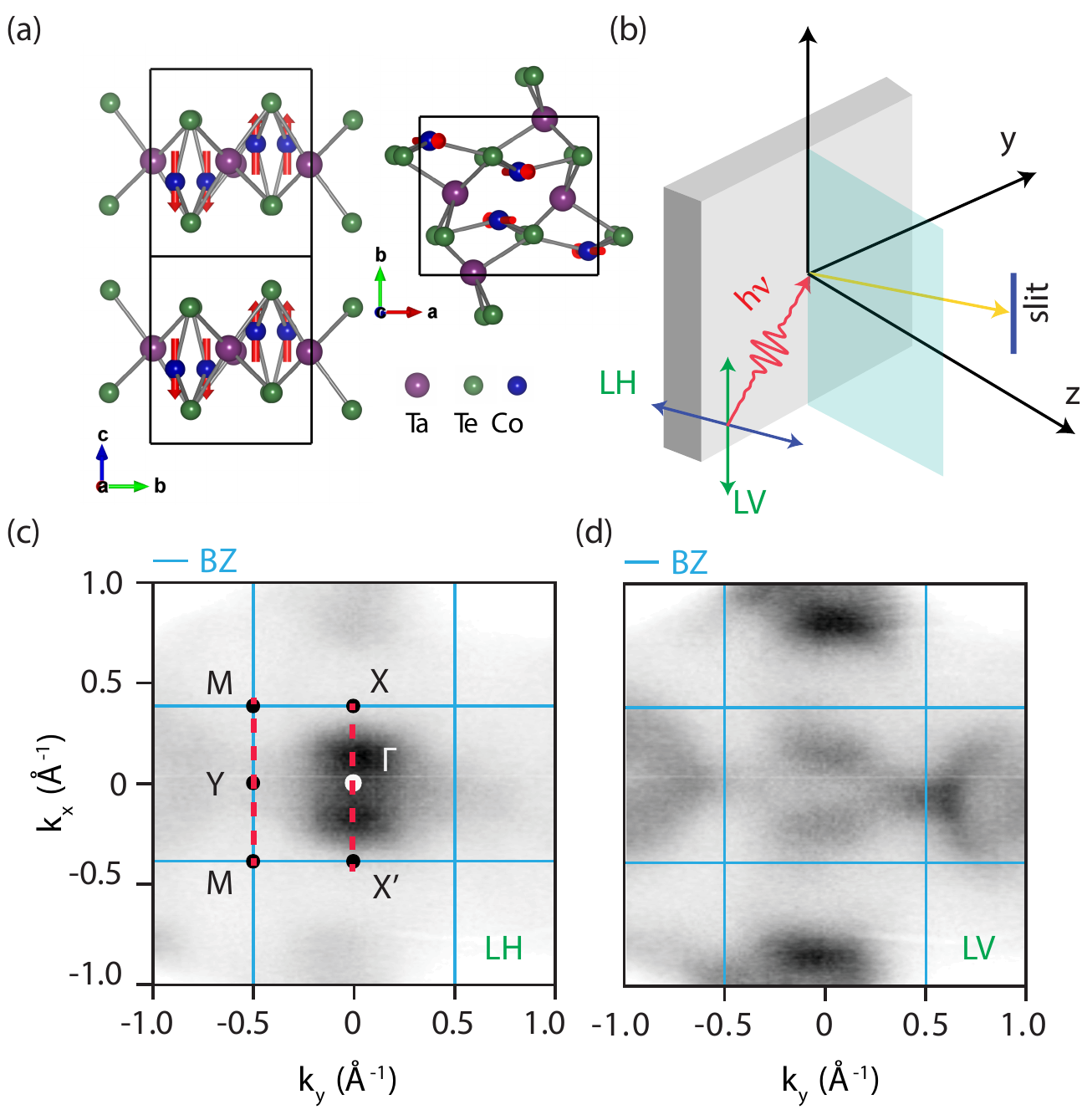}
    \caption{(a) Top and side view of the TaCoTe$_2$ crystal structure in the AFM$_z$ configuration. Red arrows indicate the magnetic moments on the Co atoms. The black solid box identifies the unit cell. (b) The geometry of the experimental setup showing the horizontal (LH) and vertical (LV) light polarization vector. Fermi surface map collected within 20 meV of the Fermi energy with (c) LH and (d) LV polarization in ARPES at 77 K, with the BZ marked with blue lines. Fermi surface maps reflect the monoclinic crystal structure of TaCoTe$_2$, with a different elongation along the $k_x$ and $k_y$ axis.}
    \label{fig1}
\end{figure}

\begin{figure}
    \centering
    \includegraphics[width=0.9\textwidth]{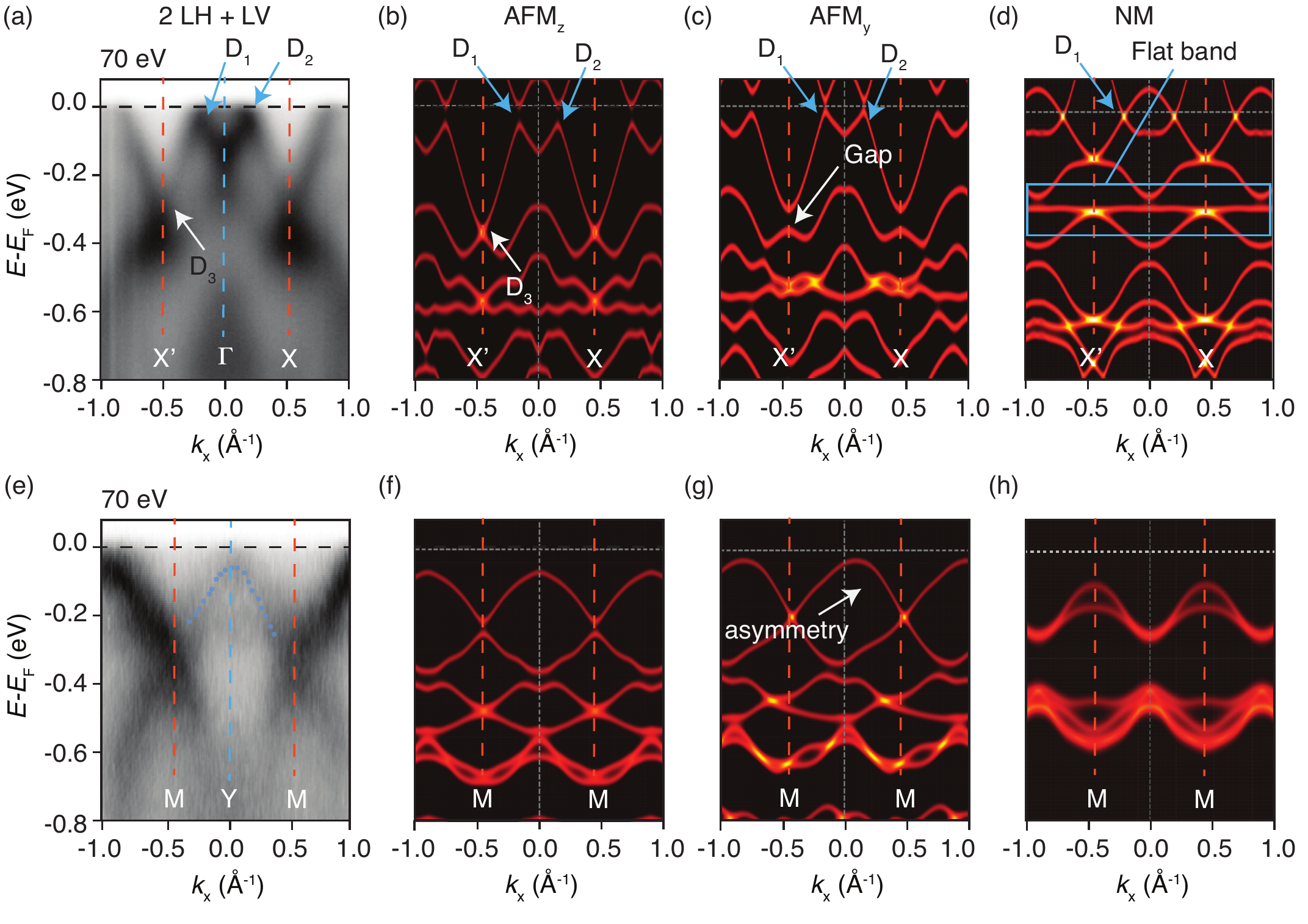}
    \caption{(a) ARPES scan along the X'-$\Gamma$-X direction for 70 eV photon energy. Calculated spectral function for monolayer TaCoTe$_2$ in the (b) AFM$_z$, (c) AFM$_y$, and the (d) NM configuration. (e) ARPES map along the M-Y-M direction collected with LV polarization to better see the BZ boundaries, and the corresponding DFT calculations for monolayer TaCoTe$_2$ in the (f) AFM$_z$, (g) AFM$_y$, and (h) NM configuration. Clearly, the theoretical results for the AFM$_z$ configuration are the closest to the measured ARPES spectrum. Note that the ARPES spectra have been shown as 2LH+LV to account for both light polarizations and also the projections of LH, which is 50$\%$ in plane and 50$\%$ out of plane. This ensures a better visibility for the spectral features which might be dependent on the light polarization.}
    \label{fig2}
\end{figure}
 
Bulk TaCoTe$_2$ has a Van der Waals type structure with individual monolayers stacked along the [001] crystallographic direction (Fig.\ref{fig1}a) with nonsymmorphic symmetry \cite{Wu_2022, Cavanagh_2022, Junzhang_2017, Schoop_2016} (see Supplementary information for details of growth and characterization). The crystal structure belongs to the monoclinic space group $P2_{1}/c$ (number 14) and gives rise to the constant energy surfaces in reciprocal space shown in Fig.\ref{fig1}c-d. Along with the constant energy ARPES maps collected at 20 meV below the Fermi energy (E$_F$), we show the Brillouin zone (BZ, blue lines) and the high-symmetry points in Fig.\ref{fig1} c. The ARPES spectra were collected with linearly polarized light, i.e. with horizontal (LH) or vertical (LV) polarization (Fig.\ref{fig1}b). In the photoelectron excitation process, the photoemission intensity strongly depends on the light polarization vector direction, in particular on its parity with respect to the system’s mirror plane. In the assumption of an even symmetry for the final state (\cite{Hermanson_1977, Damascelli_2004,Day_2019,Mazzola_20017}), an even (or odd) light polarization vector with couple with orbitals with even (odd) parity, to give overall matrix elements even under reflection with respect to the mirror plane. In our case, in a simplistic picture, this means that the light polarization vector couples with those orbitals in the system displaying a non-zero component parallel to the light vector. In particular, LV is only sensitive to the in-plane orbitals, while LH is equally sensitive to in-plane as well as  out-of-plane orbitals (given the 45 degrees of incidence angle). By using the combination of LH and LV, we conclude that the near E$_F$ electronic structure is given by an admixture of both in and out of plane orbitals, with very little variation in the spectral intensity. This allows us to better identify the spectral features across the BZ and to make a more reliable comparison with the DFT calculations.

To understand the experimental electronic structure, we collected energy vs momentum spectra by using ARPES. We first notice that the layers of bulk TaCoTe$_2$ are weakly bound together by van der Waals forces. Thus, we expect that the electronic structure is intrinsically two-dimensional, with small, if any, electronic dispersion along the $z$ direction (perpendicular to the layers). The absence of $k_z$ dispersion in TaCoTe$_2$ is confirmed by the qualitatively similar shapes of the spectra at various photon energies \cite{Damascelli_2004, Mazzola_2014} (Supplementary Fig.~S1). This behaviour is fully consistent with a two-dimensional electronic structure. The only change observed experimentally as a function of photon energy is a variation of the spectral intensity across the BZ, likely connected to the photoelectron matrix elements \cite{Moser_2017, Day_2019}, as also suggested by the data collected for both LH and LV polarization (Supplementary Fig.~S2). Our results demonstrate that TaCoTe$_2$ behaves electronically as a two-dimensional system, thus, without loss of generality, we will describe the electronic properties of this compound with those of a monolayer.

\begin{figure}
    \centering
    \includegraphics[width=0.47\textwidth]{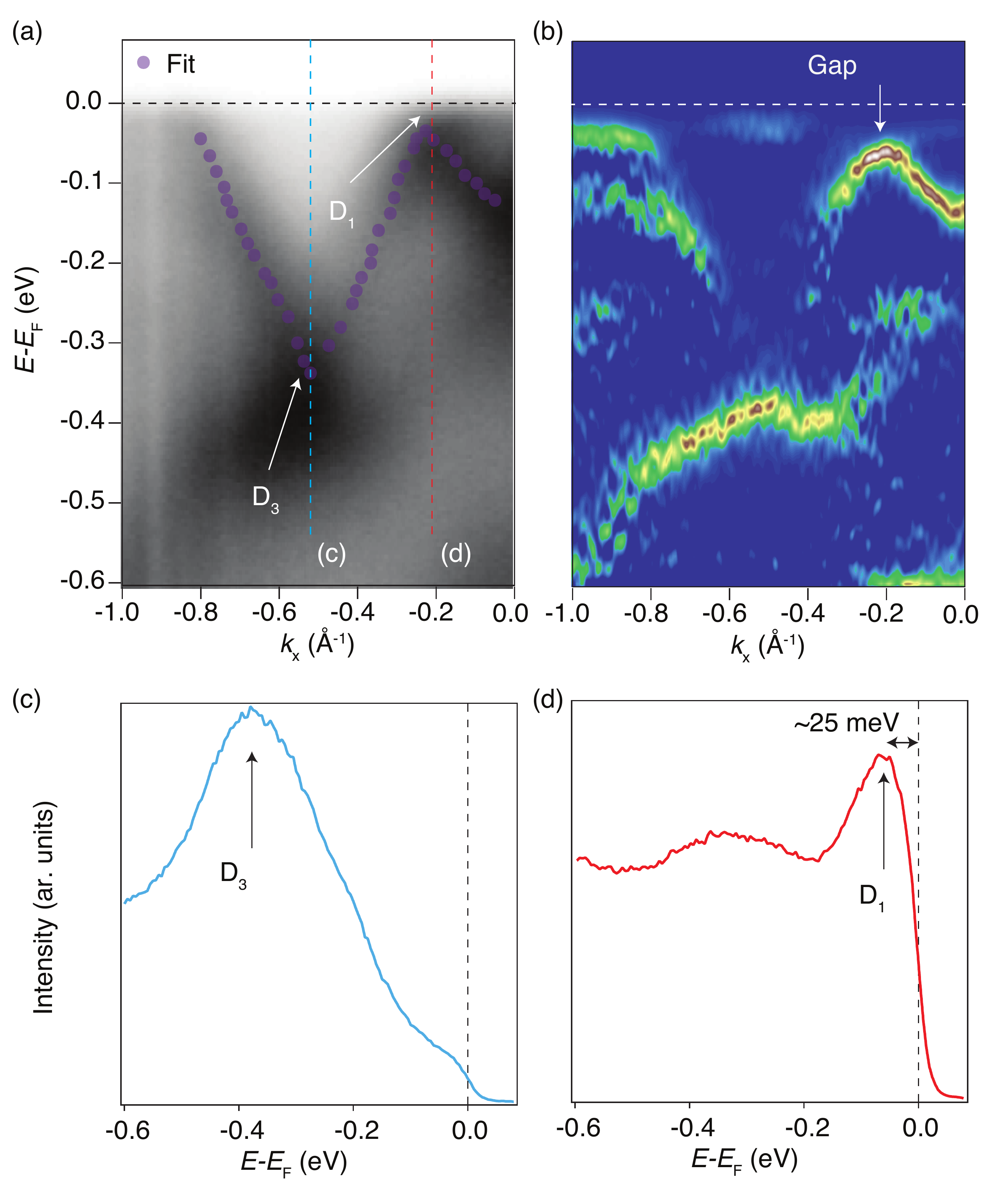}
    \caption{(a) Zoomed-in maps of the region near the Dirac dispersions D$_1$ and D$_3$. Purple dots are the positions of maximum intensity of the ARPES data, obtained by fitting the energy distribution curves (EDCs), and show the opening of a gap in D$_1$ but not in D$_3$. (b) 2nd derivative plot corresponding to the EDCs in panel (a). (c) EDC across D$_3$ at the X' point, showing the absence of a gapped two-peak structure. (d) EDC across D$_1$ showing the opening of a gap of at least 25 meV.}
    \label{fig3}
\end{figure}

From DFT, the system's lowest energy configuration is obtained for magnetic moments on the Co sites arranged in a bi-collinear type AFM order in each layer (Fig.~\ref{fig1}a). Such moments are along the $z$ direction (AFM$_z$). The energy difference between this and the AFM$_y$ order is very small, i.e. ~9 meV/ unit cell~\cite{Li_2019}, but their effect on the electronic structure is sizeable, with the realization of markedly different features identifiable in the spectra (as summarized in Fig. 2). Both AFM$_z$ and AFM$_y$ break the time-reversal, as well as the inversion symmetry of the system ($\mathcal{T}$ and $\mathcal{P}$), while preserving the combined $\mathcal{PT}$ symmetry, important for the transport and optical properties of the system \cite{Bender_1998, Zyablovsky_2014, Ganainy_2018}. The main differences between AFM$_z$ and AFM$_y$ are visible in the nonsymmorphic symmetry of the system (See Supplementary information), which manifests directly in different energy vs momentum electronic structures.

To unveil the magnetic ordering of this compound, we compared non-magnetic (NM) DFT calculations along with those for the antiferromagnetic AFM$_z$ and AFM$_y$ orders, with the measured energy-momentum spectra (Fig.~\ref{fig2}). Figs.~\ref{fig2}a-d displays the results along the high-symmetry direction X'-$\Gamma$-X, Fig.~\ref{fig2}e-h for the M-Y-M line, as shown in the BZ of Fig.~\ref{fig1}c. Evidently, all configurations share common features. Along the X'-$\Gamma$-X direction, AFM$_z$ (Fig.~\ref{fig2}b), AFM$_y$ (Fig.~\ref{fig2}c), and NM (Fig.~\ref{fig2}d) display Dirac cones located around the Fermi level (labeled D$_{1,2}$). This is also similar to the experiment of Fig.~\ref{fig2}a. However, a closer inspection reveals important differences: the NM order features a flat band centered around -0.3 eV and a cosine-like dispersion pinned to such a flat band (see the blue guide to the eye). These features are absent in the experiment, as well as in the two magnetic calculations. In addition, the NM configuration along the M-Y-M direction (Fig.~\ref{fig2}h) does not exhibit any band with a maximum at $\Gamma$ near the Fermi level. On the contrary, this feature is evident in the experimental data of Fig.~\ref{fig2}e. Based on the comparison between ARPES and DFT calculations, one can rule out the NM configuration as the experimental ground state of TaCoTe$_2$.

The differences between AFM$_z$ and AFM$_y$ along X'-$\Gamma$-X are also measured and calculated: Along the M-Y-M path, as a consequence of the magnetism along y, the AFM$_y$ order shifts (about $25\%$) the maximum of the band at Y towards the M point, creating an extremely large asymmetry in the electronic dispersion, see Fig.~\ref{fig2}g. This asymmetry, in contrast, is neither observed in the experiment nor predicted in the AFM$_z$ calculation (Fig.~\ref{fig2}f). Furthermore, the opening of a gap of a Dirac-like dispersion at -0.3 eV at the X and X' points for the AFM$_y$ configuration (indicated as D$_3$) is absent in the AFM$_z$ order and not detected by ARPES, despite its size is expected to be much larger than the experimental resolutions (12 meV and 0.018\AA$^{-1}$ for energy and momentum, respectively). This allows us to conclude that the electronic structure of TaCoTe$_2$ is consistent with a magnetic order hosting magnetic moments aligned along the z direction, i.e., AFM$_z$. This is also consistent with our magnetization measurements, which suggest the existence of an easy axis mainly along the out-of-plane direction (Supplementary Fig.~S3). Such a magnetization, shows also non-trivial magnetic signatures in the curves, i.e. saturating magnetization, which are in full agreement with what predicted in Ref.\cite{Li_2019}. In addition, as mentioned above, AFM$_z$ is the most stable configuration found for this system. We also notice that ARPES does not resolve the fine details that DFT calculations reveal for binding energies lower than $-0.4$~eV, regardless of the magnetic order. We believe that the large $k_z$ broadening of the samples, which indeed manifests as 'shadowing' of the electronic structure, combined to the strongly varying matrix elements (as shown in Supplementary information) might be the reasons behind the apparent discrepancies. It is still worth mentioning that although transport, ARPES and DFT results indicate that TaCoTe$_2$ realizes a possible AFMz order, neutron scattering experiments would be also beneficial to conclusively determine the precise magnetic ground state, but it is beyond the current scope of this work.

Together with the identification of the magnetic order in TaCoTe$_2$, our data demonstrates that SOC plays an important role in opening energy gaps at the Dirac points. This can be seen in the AFM$_z$ electronic structure calculations of Fig.~\ref{fig2}b, where D$_{1,2}$ develops energy gaps right at the Fermi energy, while no gap is observed in D$_3$. Such gaps are candidate $k$-space \textit{loci} to enable topologically non-trivial behaviour \cite{Sch_2019, Chen_2009, Xia_2009, Chagas_2022}. We also experimentally detect such gaps: we show a zoom around D$_1$ and D$_3$ (Fig.~ \ref{fig3}~a), the intensity curvature  plot in Fig.~\ref{fig3}~b \cite{Zhang_2011}, and the extracted energy-distribution curved across the X' point and exactly across D$_1$ (Fig.~\ref{fig3}~c,d). The energy profile at the X' point D$_3$ does not show any gap, but rather a single peak, consistently with the predicted antiferromagnetic behavior along the z direction (Fig.~\ref{fig3}c). As for D$_2$, the small SOC-induced gap at the Fermi level predicted by DFT ($\approx 20$~meV) is more challenging to be observed with state-of-the-art experimental apparatuses. However, by fitting the dispersion with Lorentzian curves (purple markers in Fig.~\ref{fig3}a (See Supplementary Information for details), we estimate that the top of the band is located $\sim 25\pm10$~meV below the Fermi level. We stress that a proper quantification of this gap is difficult via ARPES, because it involves unoccupied states that are not accessible to ARPES. However, the value of the gap ($\sim 25\pm10$~meV) adduced from the experimental data is a lower limit compatible with the calculated gap value. We also show in the Supplementary information how the observed peak is lower in binding energy compared to the Fermi level edge, extracted from EDCs in a region without bands, i.e. at $k_x$=$-0.5$~\AA$^{-1}$.
 
\begin{figure}
    \centering
    \includegraphics[width=0.50\textwidth]{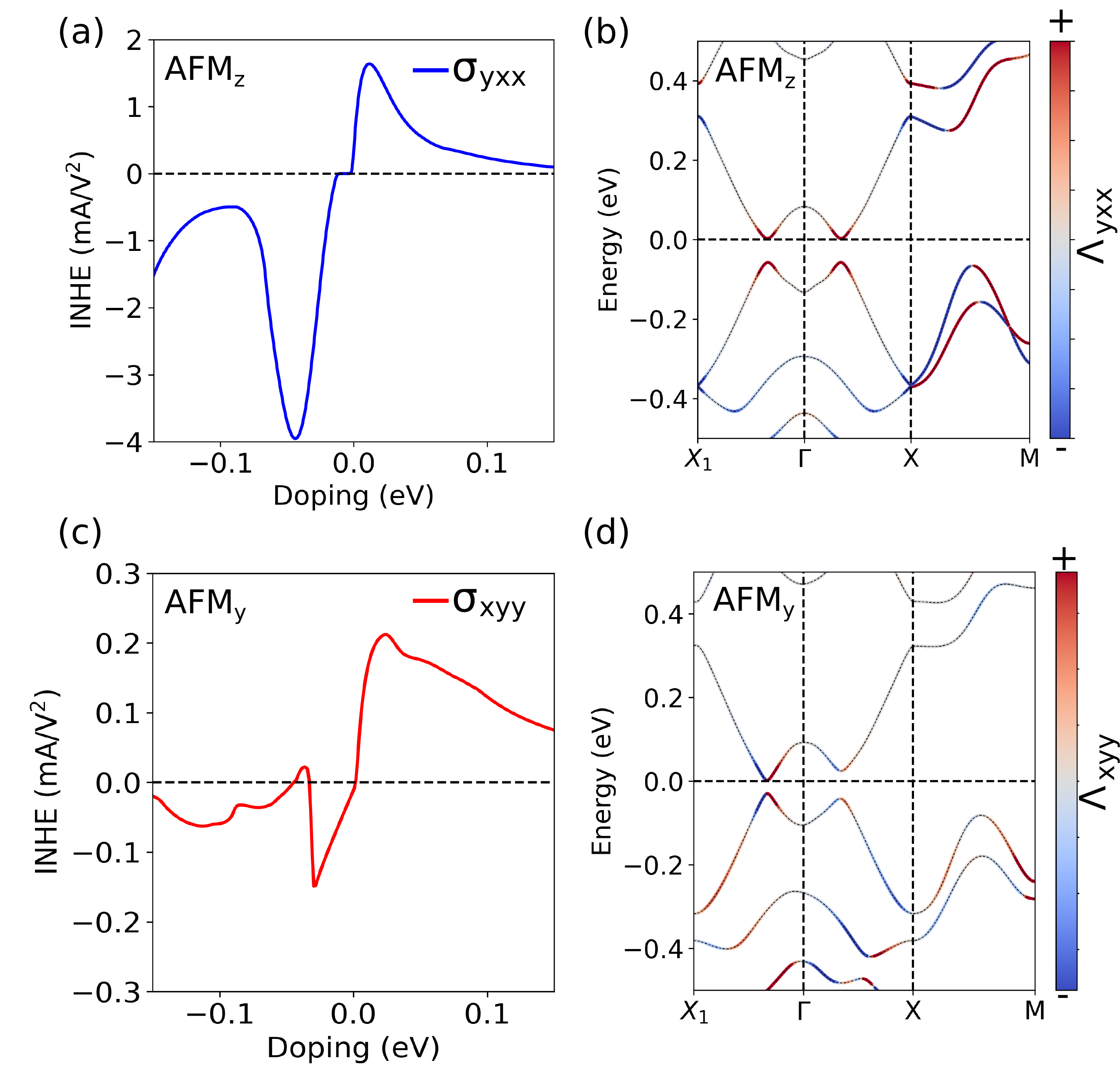}
    \caption{Intrinsic non-linear Hall conductivity (INHE) and the distribution of the $\Lambda^n_{\alpha\beta\gamma}({\bf k})$ on the band structure of TaCoTe$_2$ in the (a)-(b) AFM$_z$, and (c)-(d) AFM$_y$ phase. Note that, distinct symmetries of the two different magnetic configurations enforce different non-zero components of the INHE. Clearly, the INHE is highly sensitive to the Neel vector orientation. 
    \label{fig4}}
\end{figure}

The presence of SOC gaps is of paramount importance for giving rise to exotic transport phenomena, and in particular for the intrinsic nonlinear Hall effect (INHE) in antiferromagnetic Dirac systems. The INHE arises from the field correction to the Berry curvature in the presence of an external electric field, and it has recently gained significant attention, due to its intrinsic nature \cite{inhe2014,inhe_2,inhe_xiao,inhe_3}. Moreover, INHE has been attributed to band geometric quantities like the quantum metric tensor and it has been described as the quantum metric dipole-induced non-linear Hall effect \cite{inhe_xiao,inhe_3,Graf_2021, Gianfrate_2020, Villegas_2021} (See Supplementary information for a mathematical derivation of the INHE).

We used DFT to compute the INHE ($\sigma$) for the monolayer TaCoTe$_2$ in the AFM$_y$ and AFM$_z$ magnetic configurations. The NM state preserves both the $\mathcal{P}$ and $\mathcal{T}$ symmetry, and, as a result, all the components of the INHE vanish identically in this case. In the AFM phases, both $\mathcal{P}$ and $\mathcal{T}$ symmetries are broken, but the combined $\mathcal{PT}$ symmetry results in non-vanishing INHE. Specifically, in the AFM$_z$ phase, the relevant symmetry is $\mathcal{\tilde{C}}_{2y}$, which forces the $\sigma_{xyy}$ component to vanish, while the $\sigma_{yxx}$ is non-zero. In contrast, the AFM$_y$ structure hosts the $\tilde{M}_{y}$ symmetry, which results in a vanishing $\sigma_{yxx}$ and a non-zero $\sigma_{xyy}$. The INHE is shown in Fig.~\ref{fig4}a for the AFM$_z$ configuration, and in Fig.~\ref{fig4}c for the AFM$_y$ configuration. In both cases, the INHE vanishes identically inside the bandgap, and develops peaks near the band edge and decays rapidly when the chemical potential is shifted away. It exhibits an opposite sign for the electron and hole doping for both AFM$_z$ and AFM$_y$ order. The INHE value is found to be of the order of 1 mA/V$^2$, which is comparable to the recently reported values in metallic antiferromagnets \cite{inhe2014,inhe_2,inhe_xiao,inhe_3}. Interestingly, the INHE is almost an order of magnitude larger for AFM$_z$ compared to AFM$_y$. Finally, given the direct link between the INHE and $\Lambda$ (band resolved contribution to the INHE as defined in Eq.4 of the Supplementary information), we present the distribution of the latter on the band structure in Figs.~\ref{fig4}b and ~\ref{fig4}d. As expected, near the band edge, $\Lambda^n_{\alpha\beta\gamma}(k)$ has the maximum value, and it decays away from the band edge.

In conclusion, we demonstrate that TaCoTe$_2$ is an antiferromagnetic Dirac system, which hosts SOC-driven bandgaps at the Fermi level. The combination of SOC effects, magnetism, and time-reversal symmetry breaking is found to generate a non-vanishing INHE, which influences to the transport properties of the system. The INHE in TaCoTe$_2$ is highly sensitive to the direction of the Néel vector of the AFM order, opening a novel pathway for using this compound in dissipationless electronics and spintronics. Our study indicates that TaCoTe$_2$ would provide a promising new materials platform for exploring the interplay of Dirac fermiology, SOC, magnetism, and topology.

\section{Acknowledgements}
The experiments were performed at the NFFA APE-LE beamline on the Elettra synchrotron radiation source, supported by the NFFA international facility of MUR-Italy. The work at Northeastern University was supported by the Air Force Office of Scientific Research under Award No. FA9550-20-1-0322, and benefited from the computational resources of Northeastern University’s Advanced Scientific Computation Center (ASCC) and the Discovery Cluster. AA acknowledges the Science and Engineering Research Board (SERB) and the Department of Science and Technology (DST) of the Government of India for financial support. DF\ acknowledges financial support from the Spanish Ministry of Economy and Competitiveness, through the ``Mar\'{\i}a de Maeztu" Programme for Units of Excellence in R\&D (CEX2018--000805--M) and project  PID2019--109525RB--I00. JH acknowledges the support by the U.S. Department of Energy, Office of Science, Office of Basic Energy Sciences, under Award No. DE-SC0022006 (crystal growth, electronic and magnetic property measurements). F.M. greatly acknowledges the SoE action of pnrr, number SOE\_0000068..
%%%%%%%%%%%%%%%%%%%%%%%%%%%%%%%%%%%%%%%%%%%%%%%%%%%%%%%%%%%%%%%%%%%%%
%% The same is true for Supporting Information, which should use the
%% suppinfo environment.
%%%%%%%%%%%%%%%%%%%%%%%%%%%%%%%%%%%%%%%%%%%%%%%%%%%%%%%%%%%%%%%%%%%%%
\section*{Authors contributions}
FM and BG contributed equally to this work. F.M, I.V, J.F., D.M., G.R, D. F. and A.P carried out the ARPES measurements and analyzed the data. B.G. performed the theoretical calculations under the guidance of A.B. and A.A. J.H. and G.A. grew the samples and characterized it. All authors wrote the manuscript and contributed to the scientific discussion.

%\begin{suppinfo}
%This will usually read something like: ``Experimental procedures and
%characterization data for all new compounds. The class will
%automatically add a sentence pointing to the information on-line:
%\end{suppinfo}
%%%%%%%%%%%%%%%%%%%%%%%%%%%%%%%%%%%%%%%%%%%%%%%%%%%%%%%%%%%%%%%%%%%%%
%% The appropriate \bibliography command should be placed here.
%% Notice that the class file automatically sets \bibliographystyle
%% and also names the section correctly.
%%%%%%%%%%%%%%%%%%%%%%%%%%%%%%%%%%%%%%%%%%%%%%%%%%%%%%%%%%%%%%%%%%%%%

\section{AFM$_z$ and AFM$_y$ differences}
The main differences between the AFM$_z$ and AFM$_y$ orders are expected due to the nonsymmorphic symmetries, which are characterized by the combination of $\mathcal{P}$, a two-fold screw rotation symmetry along the y-axis~$\mathcal{\tilde{C}}_{2y}=\{\mathcal{C}_{2y}| \frac{1}{2} \frac{1}{2} 0\}$, and a glide mirror symmetry perpendicular to the y-axis $\mathcal{\tilde{M}}_y=\{\mathcal{M}_{y}|\frac{1}{2} \frac{1}{2} 0\}$. In particular, the AFM$_z$ order preserves the $\mathcal{\tilde{C}}_{2y}$, while the AFM$_y$ order hosts $\mathcal{\tilde{M}}_{y}$. This difference is not only important for transport properties, but it is responsible for giving rise to significantly different $k$-resolved electronic dispersion. 

\newpage

\section{Photon energy ARPES and light polarization dependence}

\begin{figure*}[h!]
    \centering
    \includegraphics[width=0.9\textwidth]{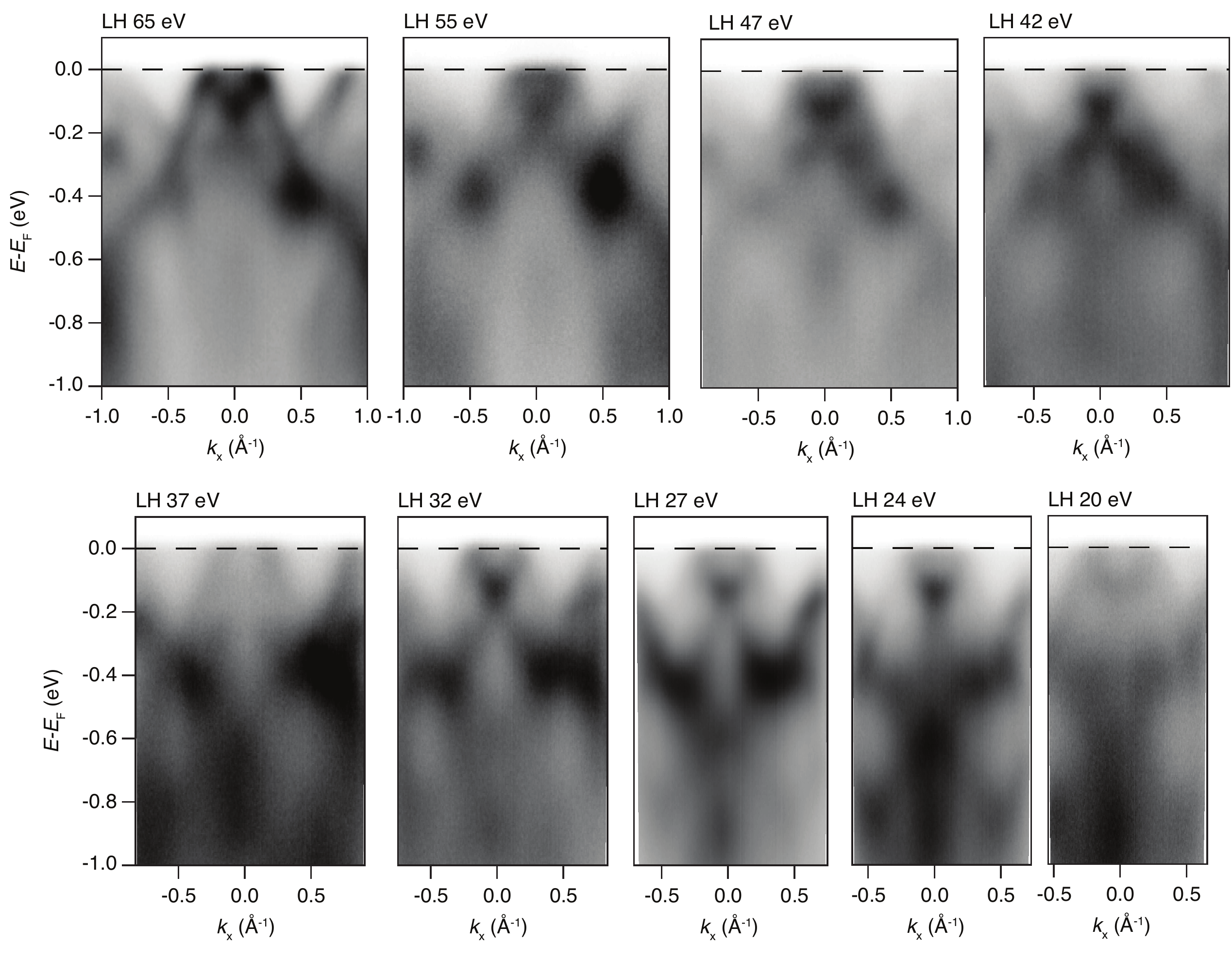}
    \caption{ARPES energy-momentum spectra collected wit linear horizontal light polarization and for different photon energies, as indicated in the panels. The main difference is a shift of the spectral weight in the spectra, but no significant change in the electronic structure can be observed. This confirms the lack of $k_z$ dispersion, thus a two dimensional character for the electronic structure of TaCoTe$_2$.}
    \label{S1}
\end{figure*}

\clearpage

\begin{figure*}[h!]
    \centering
    \includegraphics[width=0.8\textwidth]{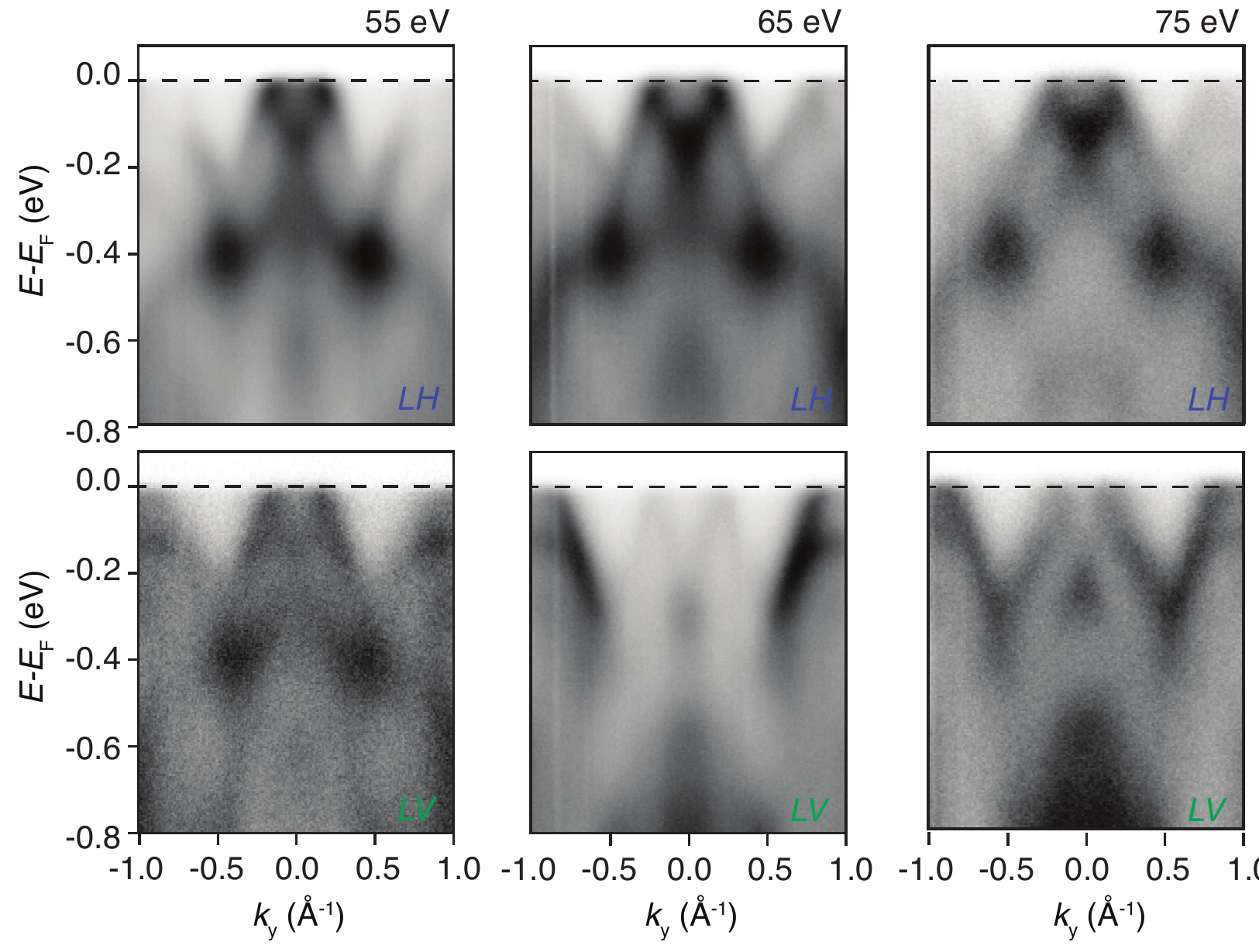}
    \caption{ARPES energy-momentum spectra at selected photon energies (where the intensity was prominent) and for both LH and LV. The main differences can be brought back to a redistribution of the spectral weight, indicating a mixed orbital character for the electronic structure collected within the energy region presented.}
    \label{S2}
\end{figure*}

\section{Sample growth and characterization}
TaCoTe$_2$ single crystals were grown by a chemical vapor transport method . The stoichiometric mixture of Ta, Co, and Te powders was sealed in a quartz tube with TeCl$_4$ being used as transport agent. Thin plate-like single crystals with metallic luster can be obtained via the chemical vapor growth with a temperature gradient of 900$^\circ$C - 750$^\circ$C. The composition and structure of the crystals were checked by Energy-dispersive x-ray spectrometer and x-ray diffractometer respectively.

\section{Methods and samples preparations}
The samples were cleaved in ultrahigh vacuum at the base pressure of $1\times10^{-10}$~mbar. The ARPES measurements were performed at the NFFA APE-Low Energy  beamline, at $77$~K, by using a Scienta DA30 hemispherical analyzer with energy and momentum resolutions better than 12~meV and 0.02~\AA$^{-1}$, respectively.

\section{ARPES Fitting Details}
The electronic structure of Fig.3a of the main text has been fitted by using Lorentzian curves convoluted by a Gaussian, which accounts for the energy resolution of the instrument. For performing the fit, the ARPES spectra have been decomposed in energy distribution curves (EDCs) and each EDC has been fitted the way here described.

\section{Details of the {\it{ab-initio}} calculations}
The density functional theory based $\it{ab-initio}$ calculations were performed using the generalized gradient approximation framework (GGA-PBE) as implemented in VASP code ~\cite{Kohn_dft,Perdew_gga,vasp,vasp_paw}. A k-grid of 12$\times$12$\times$8 was used for the BZ integration. The kinetic energy cutoff for the plane wave basis was set to 400 eV. The Wannier function-based tight binding modeling is done by considering the s and d orbitals of Ta, the p orbital of  Te, and the d orbitals of Co ~\cite{wannier,wannierv3}.

\clearpage

\section{Temperature Dependent Resistivity and Magnetization Data}

\begin{figure}[h!]
    \centering
    \includegraphics[width=0.85\textwidth]{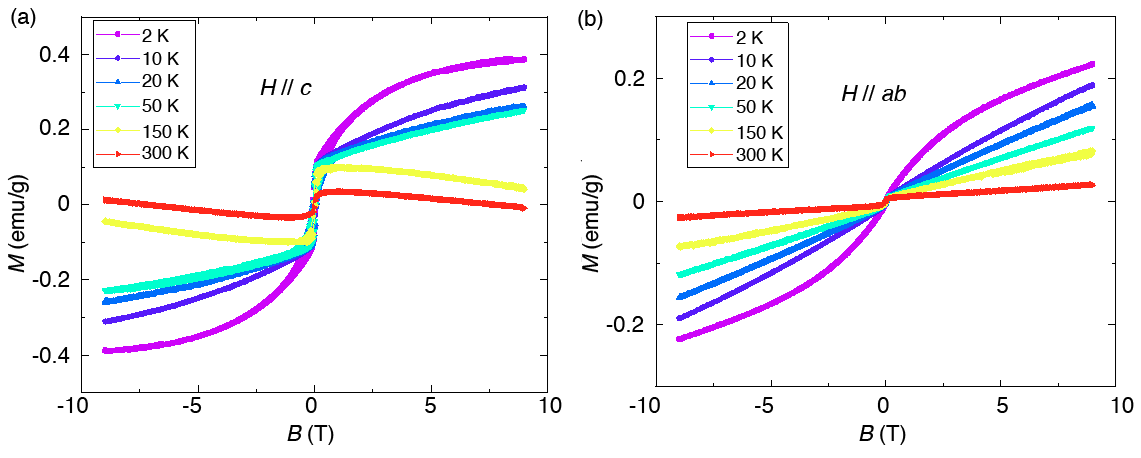}
    \caption{Temperature dependent magnetization data for TaCoTe$_2$ for the applied magnetic field along (a) $B||c$ direction, and (b) $B||ab$ direction. Clearly, the isothermal magnetization displays a typical moment polarization behavior for all measured temperatures up to 300 K, under both out-of-plane ($B||c$) and in-plane ($B||ab$) magnetic field orientations. Such behavior signatures the presence of magnetic order in the system. The out-of-plane magnetization is roughly twice as large as the in-plane one, suggesting that the easy axis is mainly along the out-of-plane direction. }
    %ferromagnetic, ferrimagnetic, or antiferromagnetic order with canted moment. }
    \label{St}
\end{figure}

\begin{figure}[h!]
    \centering
    \includegraphics[width=0.45\textwidth]{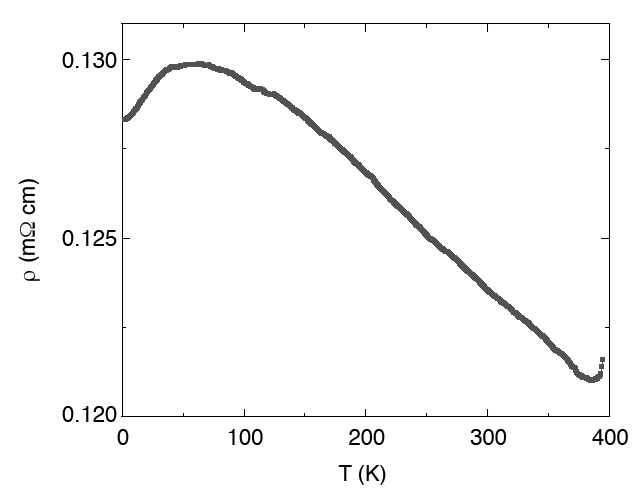}
    \caption{Temperature dependent resistivity of TaCoTe$_2$. The resistivity does not strongly vary with the temperature. Overall, it displays an non-metallic transport with resistivity increases upon cooling, followed by a resistivity down turn around 50 K. A transition-like behavior characterized by resistivity upturn is observed around 380K, which is reproducible in multiple samples with slight variation in upturn temperature.}
    \label{Srho}
\end{figure}

\section{INHE and Berry curvature connection}
In a simple semi-classical model, INHE can be understood from the presence of a finite Berry connection polarizability (BCP) in the system. The BCP for the the n'th band can be expressed as,

\begin{equation}
G^{n}_{ab}({\bf k})= 2 {\text{Re}} \sum_{m\neq n} \frac{A_a^{nm}({\bf k})A_b^{mn}(\bf k)}{(\epsilon_n({\bf k})-\epsilon_m({\bf k}))},
\end{equation}
where, $A^{nm}_a=\langle u_n \vline i\partial_a \vline u_m \rangle$ is the interband Berry connection, and $\vline u_n \rangle$ are the unperturbed Bloch states with band energies $\epsilon_n$. In the presence of an external electric field ($E_b$), the net generated Berry connection can be obtained as, 
\begin{equation}
A^E_a({\bf k})= G_{ab}({\bf k})E_b
\end{equation}
The field-induced Berry curvature $\Omega_E(k)=\nabla \times A^E(k)$ results in a non-linear Hall-like response. The most important feature of this response is that it is intrinsic, and unlike the Berry curvature dipole induced nonlinear Hall conductivity, INHE does not depend on the relaxation time ($\tau$). The final expression for the INHE is given by, 

\begin{equation}
\sigma_{\alpha\beta\gamma}= \int \frac{d^3k}{(2\pi)^3} \sum_n \Lambda^n_{\alpha\beta\gamma}({\bf k}) \frac{df(\epsilon_n({\bf k}))}{d\epsilon_n({\bf k})}~.
\label{eq_inhe_full}
\end{equation}
where, $f(\epsilon_n)$ is the Fermi Dirac distribution function. The momentum-resolved $\Lambda^n_{\alpha\beta\gamma}$ (we suppress the argument ${\bf k}$ for brevity) is expressed in terms of the interband Berry connection and the velocity operator ($v_n$) as,

\begin{equation}
%\begin{split}
\Lambda^n_{\alpha\beta\gamma}=2e^3 \sum_{m}^{\epsilon_n\neq\epsilon_m}  \Re[ \frac{v_n^\alpha A_{nm}^\beta A_{mn}^\gamma }{(\epsilon_n -\epsilon_m)}- \\
\frac{v_n^\beta A_{nm}^\alpha A_{mn}^\gamma }{(\epsilon_n -\epsilon_m)}]~.
%\end{split}
\label{eq_inhe_res}
\end{equation}

\section{Fermi level and observed gap}
The observed peak in the electronic structure, which forms the top of the bands which develop a gap, is lower in binding energy compared to the Fermi level edge, extracted from EDCs in a region without bands, i.e. at $k_x$=$-0.5$~\AA$^{-1}$. This is visible in the figure below.

\begin{figure}[h!]
    \centering
    \includegraphics[width=0.35\textwidth]{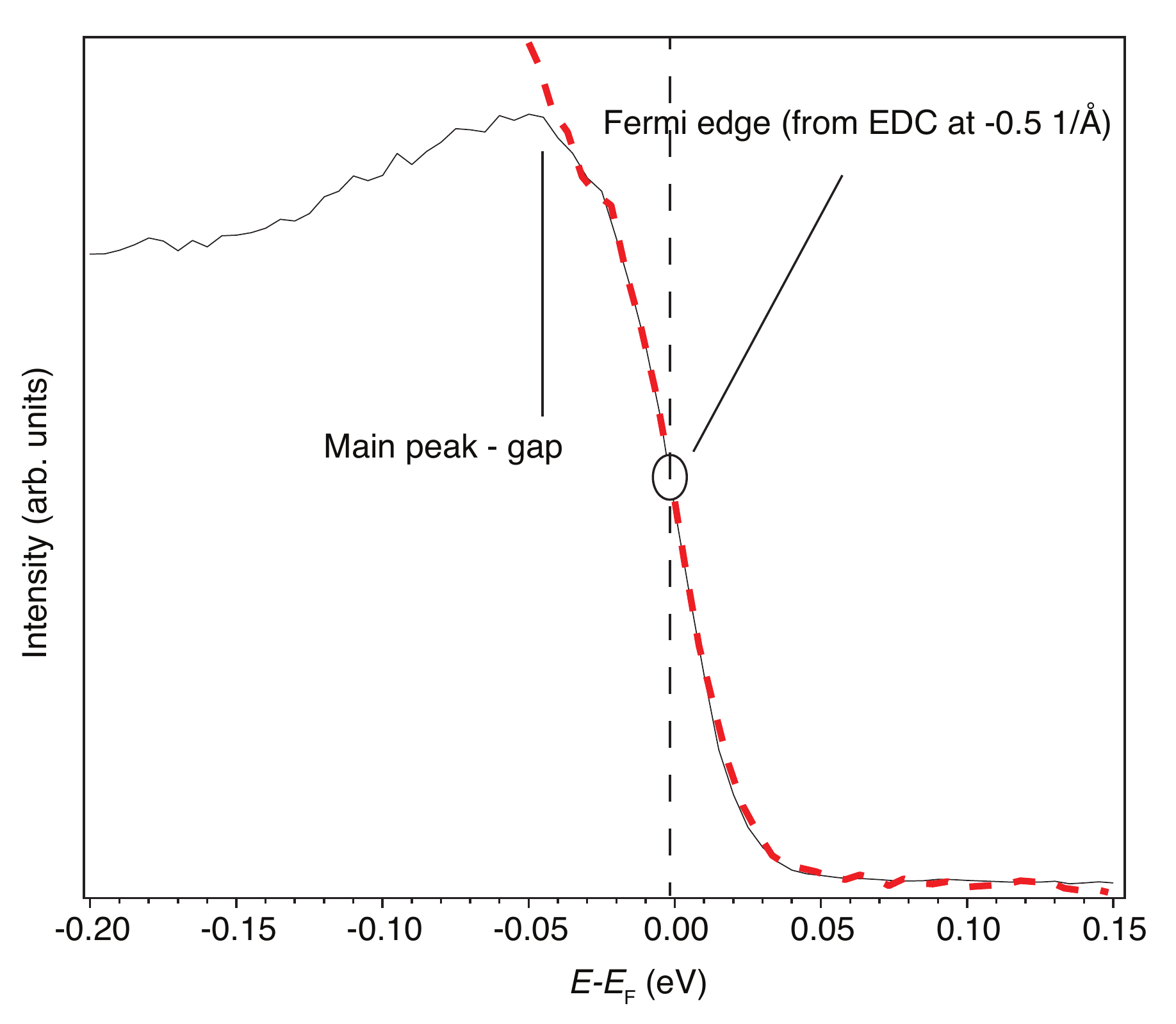}
    \caption{The red dashed line has been extracted from EDc around $k_x$=$-0.5$~\AA$^{-1}$, where there are no bands in the ARPES data. This allows us to determine the leading edge of the Fermi level as a calibration. We observe that the peak of the electronic structure discussed in the main text, formed by the band which delevlops a gap, is shifted away from the edge. The latter also, despite the increase (as observed in the figure and caused by the presence of additional bands at lower binding energies) is still visible.}
    \label{gap}
\end{figure}

\bibliography{biblio.bib}

\end{document}